# Emergent electric field induced by current-driven domain wall motion in a room-temperature antiferromagnet FeSn$_2$

Tomoyuki Yokouchi[1†], Yuta Yamane[2,3], Yasufumi Araki[4], Jun'ichi Ieda[4] and Yuki Shiomi[1]

[1]*Department of Basic Science, The University of Tokyo, Tokyo, Japan*
[2]*Frontier Research Institute for Interdisciplinary Sciences, Tohoku University, Sendai, Japan*
[3]*Research Institute of Electrical Communication, Tohoku University, Sendai, Japan*
[4]*Advanced Science Research Center, Japan Atomic Energy Agency, Tokai, Japan*

† To whom correspondence should be addressed. E-mail: yokouchi@g.ecc.u-tokyo.ac.jp

**Abstract**

**Antiferromagnets have attracted extensive interest as platforms for nanoscale spintronic devices owing to ultrafast spin dynamics and lack of a stray field. One of the crucial missing pieces in antiferromagnets is a quantum-mechanical electric field known as an emergent electric field, which has been observed for the motion of ferromagnetic spin texture. Since this phenomenon allows for the development of novel spintronic devices such as quantum inductors, its identification in antiferromagnets is vital for developing nanoscale spintronic devices. Here, we demonstrate that the motion of antiferromagnetic spin textures generates an emergent electric field. In a room-temperature antiferromagnet $FeSn_2$, we observed large current-nonlinear responses in the imaginary part of the complex impedance at and well above room temperature. This signal is attributed to an emergent electric field resulting from the nonadiabatic electron spin dynamics during the current-induced motion of antiferromagnetic domain walls. Notably, the observed electric response is strongly enhanced as the sample size decreases and robust against magnetic fields. Our finding may pave the way for novel nanoscale quantum spintronic devices.**

Main text

Electric control of spin structures is at the heart of spintronics. In the past, the coupling between Oersted fields and the macroscopic magnetic moment of ferromagnetic domains was often used to manipulate ferromagnetic domains and domain walls (DWs). More recently, charge currents have been utilized for this purpose. In these methods, spin dynamics are excited by the transfer of angular momentum via spin-polarized conduction electrons (spin-transfer torque)[1,2] or by spin torques originating from spin-polarized electrons owing to spin-orbit coupling (spin-orbit torques)[3]. In contrast, controlling antiferromagnetic (AFM) domains and DWs has traditionally been considered challenging owing to the lack of macroscopic magnetic moment. In practice, an extremely large magnetic field is required to control AFM domains. However, recent findings have revealed that charge current can also be exploited to manipulate AFM structures via spin-transfer torque[4,5] and spin-orbit torques[6,7,8,9,10,11]. Because AFM states do not produce stray fields and their dynamics can be much faster than those of ferromagnets, antiferromagnets are considered as promising materials for application to the nanoscale spintronic devices[12,13].

Conversely, the dynamics of spin structures modulates the electron motions, generating the so-called emergent electric field that contributes to a voltage called spin motive force[14,15,16,17]. In ferromagnetic spin textures, the emergent electric field is given by

$$e_i = \frac{\hbar}{2e}\left[\mathbf{m}\cdot\left(\frac{\partial \mathbf{m}}{\partial t}\times\frac{\partial \mathbf{m}}{\partial i}\right) + \beta\frac{\partial \mathbf{m}}{\partial t}\cdot\frac{\partial \mathbf{m}}{\partial i}\right], \quad (1)$$

where $i$ denotes the spatial coordinate ($x$, $y$, and $z$), $\mathbf{m}$ is a unit vector parallel to the direction of localized spins, $e > 0$ is the elementary charge, and $\beta$ is a dimensionless constant characterizing the nonadiabatic electron spin dynamics[14,15,16,18,19]. The first term is associated with the electron-spin Berry phase, and the second term represents nonadiabatic correction. So far, the emergent electric field have been observed in various ferromagnetic spin textures such as ferromagnetic domain walls (DWs)[20,21], single domains[22], magnetic vortecies[23], helical spin strctures[24,25,26,27], and skyrmions[28] when they are in motion. The emergent electric field was first observed when a magnetic field drives ferromagnetic DWs[20,21]. More recently, it was theoretically found that when a current drives noncollinear spin structures such as DWs and helical spin structures, the emergent electric field also appears[29]. Experimentally, the emergent electric field due to the current-driven motion of helical spin structures has been investigated[24,26,27]. Incidentally, because this emergent phenomenon is akin to the inductive voltage, it is termed "emergent inductance"[29]. In particular, the magnitude of the emergent inductance increases with the decreasing cross-sectional area of the samples, as opposed to the conventional inductors[29]. Thus, the emergent induction may enable miniaturization of inductors. However, the emergent electric field observed in ferromagnetic spin textures is sensitive to external magnetic fields[24,26,27], and mutual interference caused by stray fields among nearby ferromagnetic devices may hinder high-level integration in circuits.

In this work, we focus on an emergent electric field in antiferromagnetic spin textures because the absence of macroscopic magnetic moment and stray fields in AFM states could solve these issues. While the emergent electric field has not been experimentally explored in antiferromagnetic spin textures, theoretical studies suggest the equivalent effects resulting also from AFM DW motion[30,31,32,33]. In particular, since AFM DWs can be driven by current via the spin-transfer torque, an emergent electric field accompanied by the current-driven motion of AFM DWs should be observed. First, to test this working hypothesis, we derive emergent electric field arising from the current-induced motion of AFM DWs in a two-sublattice antiferromagnet. In this case, the net emergent electric field **e** is even with respect to the Néel vector **l** because of the sublattice symmetry. The lowest-order form of **e** that depends on the spatial and temporal derivatives of **l** is thus written as

$$e_i = \frac{\mu \hbar}{2e} \frac{\partial \mathbf{l}}{\partial t} \cdot \frac{\partial \mathbf{l}}{\partial i}, \qquad (2)$$

where $\mu$ is a phenomenological dimensionless constant[30]. Here, $\mu$ is given by $\mu = p\beta$, with $p$ denoting the spin polarization of the conduction electrons in each sublattice, when the exchange coupling energy between the conduction electron spin and the local magnetizations is dominant over the inter-sublattice electron hopping energy. Here, we assume that magnetization canting accompanying AFM DW motion is negligibly small. Importantly, Eq.(2) is proportional to $\beta$ and governed by the nonadiabatic spin dynamics, in contrast to its ferromagnetic counterpart [Eq. (1)]. For a one-dimensional DW, the resultant voltage across the DW is given by

$$V_e(t) = -\frac{\hbar p\beta}{e\lambda}\frac{dX(t)}{dt}, \qquad (3)$$

where $X(t)$ is the collective coordinate representing the DW position and $\lambda$ is the DW width (Fig. 1a). In particular, when DW is driven by a time-varying current $I(t)$, $X(t)$ depends on $I(t')$ ($t'<t$) (i.e., the history of the applied current). In other words, $X(t)$ is a functional of $I(t')$. From this dependence, when a sinusoidal current $I = I_0\sin\omega t$ is input, $V_e(t)$ exhibits both current-nonlinear in-phase (Re$V$) and out-of-phase (Im$V$) voltage components. Detailed features of Re$V$ and Im$V$ depend on the pinning potential for the DW (see Supplementary Text for details and for discussion in the case of a specific pinning potential shape). In Fig. 1c, we show the calculated fundamental frequency components of the in-phase (Re$V^{1f}$) and out-of-phase (Im$V^{1f}$) parts of $V_e$ as a function of the magnitude of the current density, in which an anharmonic pinning potential is assumed (the inset of Fig. 1c and see Supplementary Text for details). Both Re$V^{1f}$ and Im$V^{1f}$ nonlinearly increase with increasing current. We note that multiple DWs contribute to the voltage output in an additive manner (Fig. 1b) because $dX/dt$ does not explicitly depend on the detailed DW profile.

We now verify this theoretical prediction experimentally in a room-temperature antiferromagnet FeSn$_2$. FeSn$_2$ has a tetragonal structure with the space group $I4/mcm$[34,35,36]. FeSn$_2$ exhibits two magnetic phase transitions: a paramagnetic-to-collinear AFM transition at $T_{N1} = 378\pm2$ K and a collinear AFM-to-canted AFM transition at $T_{N2} = 93\pm1$ K. In the collinear AFM phase, Fe spins are antiferromagnetically coupled in the (001) plane and ferromagnetically coupled along the [001] direction (Fig. 2a)[35]. The Néel vector is parallel to either the [100] or [010] direction. In other words, there are two domains, and consequently,

DWs form between them as schematically shown in Fig. 2b. Because these DWs can be driven by current through spin-transfer torque[4,5], FeSn$_2$ is a suitable material for exploring an emergent electric field accompanied by AFM DW motion. Incidentally, in the canted AFM phase, the Fe spins rotate within the (001) plane and remain ferromagnetically coupled along the [001] direction (see the inset of Fig. 2d). Hence, the domain structure does not change substantially.

In Fig. 2d, we show the temperature ($T$) dependence of magnetization ($M$). Obvious changes in the magnetization are observed at 381 K and 92 K, which are assigned to $T_{N1}$ and $T_{N2}$, respectively. Figure 2e shows the temperature dependence of $M$ for the different crystalline orientations. The magnitudes of $M$ with magnetic fields $H$ // [100] and [010] are almost the same. This result ensures the formation of magnetic domains as illustrated in Fig. 2b; since the temperature dependencies of $M$ with $H$ // [100] and [010] are identical, [100] and [010] are macroscopically equivalent. This indicates the formation of two domains with Néel vectors parallel to [100] or [010].

To investigate the emergent electric field, we mainly focus on the out-of-phase voltage component because it is difficult to decouple the in-phase voltage component of $V_e$ from the voltage arising from the resistance. We input a sinusoidal current $I = I_0\sin\omega t$ parallel to the $c$-axis and monitored the voltages using a lock-in amplifier and a four-terminal method (Fig. 1d). Figures 3a and b show the temperature dependence of the in-phase (Re$V^{1f}$) and out-of-phase (Im$V^{1f}$) components of fundamental-frequency voltage normalized by the current, namely, the real and imaginary parts of the complex impedance (Re$Z$ = Re$V^{1f}/I_0$ and Im$Z$ = Im$V^{1f}/I_0$) measured at various current densities in a microscale as-grown FeSn$_2$ single crystal (Fig. 2c and the inset of Fig. 3c). The temperature dependence of Re$Z$, which is dominated by the resistance, exhibits metallic behavior and is almost independent of the current density. In contrast, Im$Z$ highly depends on the current density in the collinear AFM phase (Fig. 3b); while Im$Z$ is almost zero for small current densities, Im$Z$ substantially increases just below $T_{N1}$ for high current densities. This feature can also be seen from the current-density dependence of Im$Z$ (Fig. 3d). Furthermore, the observed nonlinear property is also confirmed by a higher harmonic measurement (Fig. 3c); the imaginary part of the third-harmonic complex impedance (Im$Z^{3f}$) is strongly enhanced at $T_{N1}$. We note that the observed signal cannot be explained by stray capacitance; stray capacitance between voltage electrodes may induce a negative Im$Z$[37]. The stray-capacitance-induced negative Im$Z$, however, must be linear with respect to the current and independent of the magnetic structure. These features contradict our observation since Im$Z$ exhibits current nonlinearity (Fig. 3d) and is enhanced at $T_{N1}$ (Fig. 3b).

To elucidate the origin of the observed Im$Z$, we investigate the dependence of Im$Z$ on electrode-distance ($d_{electrode}$) and cross-sectional area of samples by measuring Im$Z$ for various devices with different sizes and $d_{electrode}$ (Supplementary Figs. 3 and 4). To accurately estimate the current-nonlinear Im$Z$ component for all the samples, we define the difference between Im$Z$ measured with high and low currents as Im$\Delta Z$ (see Methods and Supplementary Fig. 3). As shown in Fig. 3e, Im$\Delta Z$ is proportional to $d_{electrode}$. This result is consistent with the expected trend of the emergent electric field originating from AFM DM motion. This is because the voltage increases with an increasing number of DWs between the electrodes (Fig. 1b) and consequently increases with $d_{electrode}$. In Fig. 3f, we plot Im$\Delta Z$ measured at $I_0 = 40$ mA as a function of the cross-sectional area of the samples ($S_c$). The magnitude of Im$\Delta Z$ clearly increases with decreasing $S_c$, which is also consistent with the expected behavior of the emergent electric field accompanied by the current-driven motion of the DWs; as shown in Eq. (3), the magnitude of the emergent electric field is proportional to the DW velocity $dX/dt$. The DW velocity increases with increasing current density, and consequently increases with decreasing $S_c$ when the magnitude of the input current remains constant.

To further examine the origin of the observed signals, we investigated the magnetic field dependence. As shown in Figs. 4a and b, Im$Z$ is insensitive to the magnetic field. This result is also consistent with the emergent electric field accompanied by AFM DW motion. This is because the AFM domains are robust against a magnetic field and the number of DWs is insensitive to the magnetic field below the spin-flip magnetic field. We note that no signature of a spin-flip transition was found in the *M-H* curves up to 7 T and in the anomalous Hall resistivity up to 9 T (Supplementary Fig. 2). Incidentally, a previous report[27] pointed out that a collinear AFM spin structure could temporarily become noncollinear owing to thermal spin fluctuations, thereby producing an emergent electric field. However, the emergent electric field induced by spin fluctuations must be reduced under magnetic fields because of the suppression of spin fluctuations. This mechanism is inconsistent with our results.

The dependence of Im$Z$ on the input current frequency ($f = \omega/2\pi$) also supports that the observed Im$Z$ results from current-induced AFM DW motion. In Fig. 4c, we present the frequency dependence of Im$Z/\omega$ = Im$Z/2\pi f$ at several temperatures. The magnitude of Im$Z/\omega$ decreases with increasing frequency. This decrease may be attributed to the spins not fully following the alternating current in the high-frequency region (see also Supplementary Text for theoretical calculation). To quantitatively discuss this experimental result, we fit the data with an extended Debye model (see the solid curves in Fig. 4c and Methods). This model phenomenologically represents the frequency dependence of the DW motion[38]. Figure 4d shows the temperature dependence of relaxation time $\tau$. As expected in thermally assisted DW motion, $\tau$ obeys the Arrhenius law $\tau = \tau_0 \exp(E/k_B T)$ (solid curve). We note that, in the vicinity of $T_{N2}$, $\tau$ deviates from the Arrhenius law probably because the spin fluctuations associated with the magnetic phase transition affect $\tau$.

In conclusion, we demonstrated that the current-driven motion of AFM DWs induces the emergent electric field at and well above room temperature. The observed emergent electric field is robust against a magnetic field and exhibits nonlinearity with respect to the input current. Based on our theoretical discussion, the experimental results indicate that the anharmonicity of the pinning potential of AFM DWs plays an important role in the generation of the current-nonlinear emergent electric field. Moreover, this emergent phenomenon shifts the phase of the input current and generates higher-harmonic components, acting like a nonlinear circuit. Furthermore, the response is strongly enhanced as the sample size decreases. Our findings may pave the way for the realization of novel AFM spintronic devices, such as nanoscale quantum nonlinear circuit elements.

## Methods

### Sample preparation

Single crystals of FeSn$_2$ were synthesised using a self-flux method. The Fe and Sn powders were loaded into a quartz tube with an Fe: Sn molar ratio of 3:97. The evacuated quartz tube was then heated to 600 °C at a rate of 15 °C/hour and kept at 600 °C for 24 hours, followed by cooling to 300 °C at a rate of 2 °C/hour. Subsequently, Sn was removed by centrifugation, followed by washing with nitric acid, distilled water, and ethanol. Needle-like crystals were obtained (Fig. 2c). The typical crystal width is 10 – 100 μm, and the typical crystal length is 300 – 2000 μm.

### Magnetization measurements

Magnetization was measured with a superconducting quantum interference device magnetometer (MPMS3, Quantum Design)

### Transport measurements

Gold wires were connected to the as-grown FeSn$_2$ single crystals using a micromanipulator (Axis Pro, Micro Support) in one of three ways: (1) Gold wires were attached using silver paste (Dupont 4922), followed by curing at room temperature. (2) Gold wires were attached with silver paste (Dupont 6838), followed by curing at 180 °C for approximately 1 hour. (3) Gold wires were welded using a micro spot welder (MW-2, Micro Support). The electrode distances and thicknesses of the samples were measured using a laser microscope (VK-9500, Keyence).

Complex impedance was measured using a standard four-terminal method. We input a sine-wave current ($I = I_0 \sin \omega t$) with a current source (6221, Keithley or 155, Lake Shore) and recorded both in-phase (Re$V^{1f}$) and out-of-phase (Im$V^{1f}$) components of the fundamental frequency voltage using a lock-in amplifier (LI5650 or LI 5660, NF Corporation). The real and imaginary parts of the linear impedance are defined as Re$Z$ = Re$V^{1f}/I_0$ and Im$Z$ = Im$V^{1f}/I_0$, respectively. For the third-harmonic measurement, we measured the out-of-phase third-harmonic voltage component (Im$V^{3f}$) and defined the third harmonic imaginary part of impedance (Im$Z^{3f}$) as Im$Z^{3f}$ = Im$V^{3f}/I_0$. No background subtraction was performed for Re$Z$, Im$Z$, or Im$Z^{3f}$.

Some samples exhibit a nonzero Im$Z$ even at a low current density, which is linear with respect to the current and insensitive to magnetic phase transitions. This is assigned to the background signals from the cables connecting the sample and equipment. To estimate the current-nonlinear Im$Z$ component accurately, we subtracted the background signals in Figs. 3e and f. We define the imaginary part of the background-subtracted impedance (Im$Z$) as Im$\Delta Z$ = Im$Z(I)$ – Im$Z$(5 mA), where Im$Z(I)$ and Im$Z$(5 mA) are Im$Z$ measured at currents of $I$ and $I$ = 5 mA, respectively. The definition of Im$\Delta Z$ is schematically shown in Supplementary Fig. 3. In the main text, Im$\Delta Z$ calculated with $I$ = 40 mA is shown.

**Analysis of frequency dependence**

The frequency dependence of Im$Z/\omega$ was fitted to the Cole-Davidson equation, which is an extended Debye equation as follows[38]:

$$\mathrm{Im}Z/\omega = \mathrm{Re}\left[A_\infty + \frac{A_0 - A_\infty}{(1 + i\omega\tau)^\beta}\right],$$

where $A_\infty$ and $A_0$ are Im$Z/\omega$ at the high- and low-frequency limits, respectively. The Cole-Davidson equation can describe the distribution of the relaxation process, and $\beta$ and $\tau$ are parameters representing the distribution of the relaxation process and the characteristic relaxation process, respectively.

**Data availability**

The datasets generated and analysed during this study, and the corresponding computer codes are available from the corresponding authors on reasonable request.

**Acknowledgements** This work was supported by JST FOREST Program, Grant Number JPMJFR203H, by JSPS KAKENHI, Grant Numbers 23H01832, 23H04582, 23K17882, 23H01828 22KK0072, 22H05138, 22H05449, 22H04464, 22K03538, 21H01794, and 19H05600, by JST TI-FRIS Fellow, and by The Mitsubishi Foundation. A part of this work was carried out by the joint research of the Cryogenic Research Center, the University of Tokyo.

**Author Contributions** Y.S grew single crystals. T.Y. conducted magnetization and transport measurements and analysed data. Y. Y, Y.A, and J.I. performed the theoretical calculation. All the authors wrote the draft, discussed the results, and commented on the manuscript.

**Additional information** Supplementary information is available in the online version of the paper. Reprints and permissions information is available online at www.nature.com/reprints. Correspondence and requests for materials should be addressed to T.Y.

**Competing financial interests** The authors declare no competing financial interests.

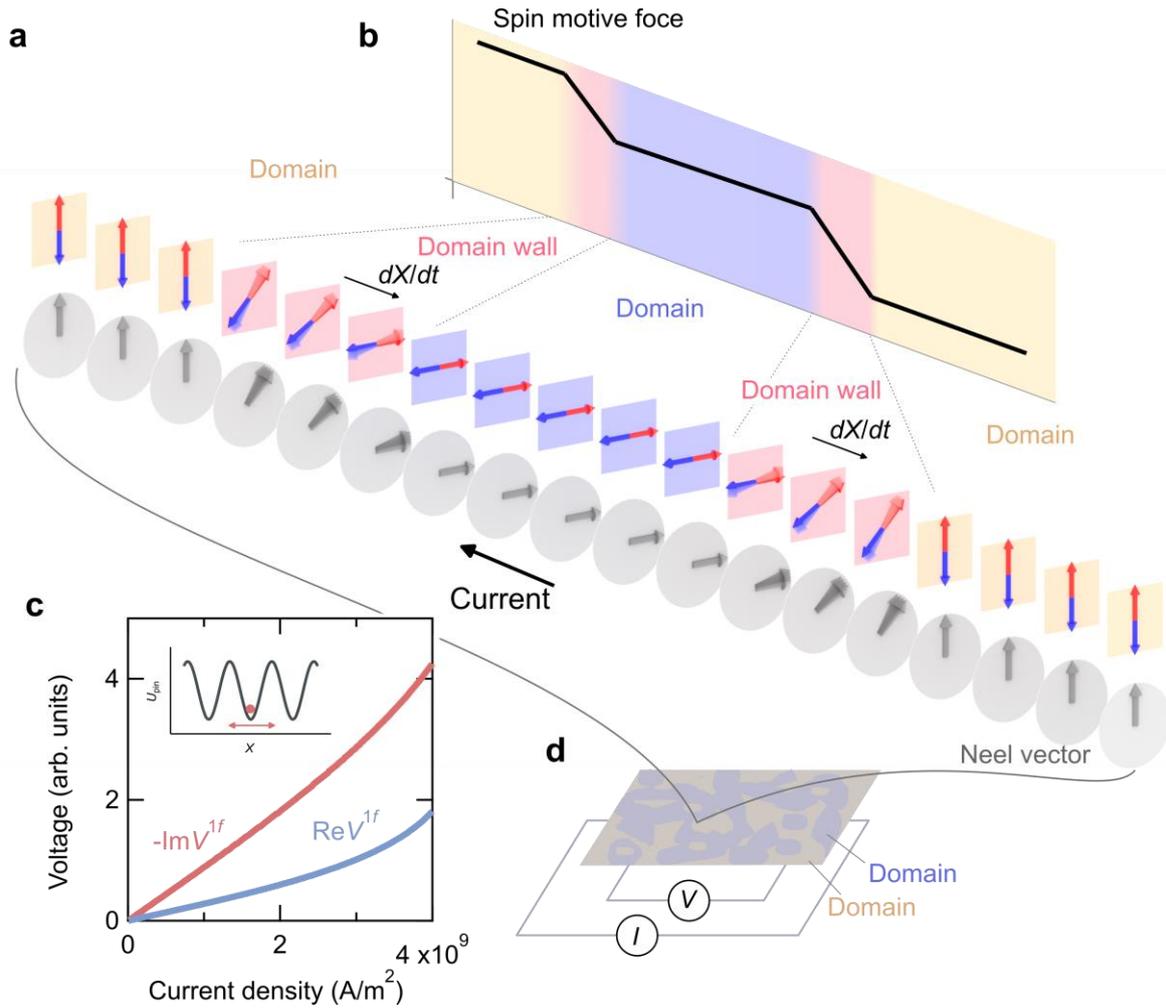

**Figure 1 | Concept of emergent electric field in AFM DW a,** A schematic for spin configuration under the current-driven motion of antiferromagnetic domain walls. Here, *X* is the collective coordinate representing the domain-wall position, and *dX/dt* is the domain-wall velocity. The red and blue arrows stand for the directions of the spins, and the gray arrows are Néel vectors. **b**, A schematic of spin motive force as a function of position. Voltage drops occurs at the domain walls. **c**, Calculated fundamental frequency components of in-phase (Re$V^{1f}$) and out-of-phase (Im$V^{1f}$) $V_e$ as a function of the magnitude of the current density. The inset schematically shows the shape of the pinning potential used in the calculation. **d**, Schematics of the experimental setup.

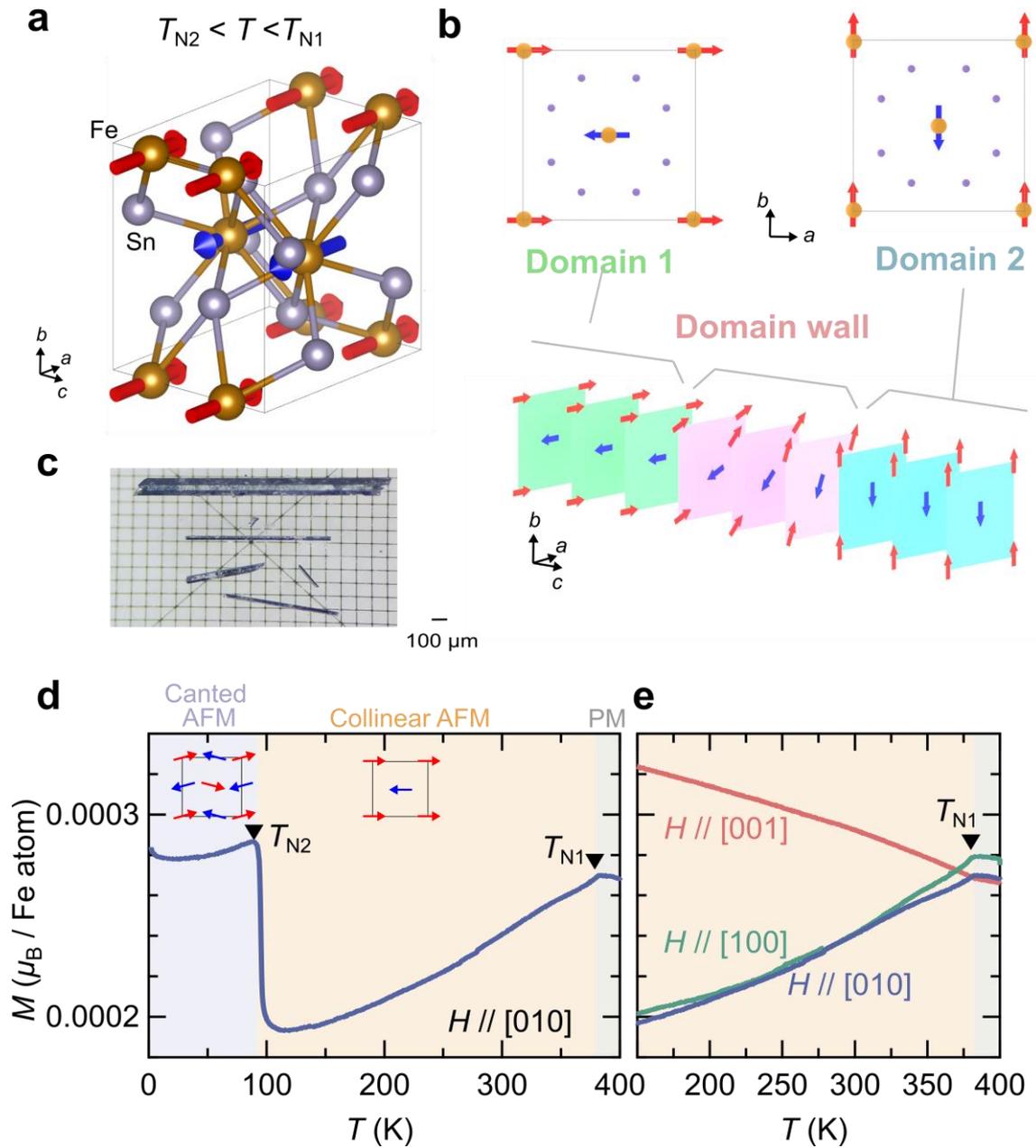

**Figure 2 | Magnetic properties of FeSn₂ a,** Crystal structure of $FeSn_2$ and AFM ordering. **b,** A schematic for magnetic domains and domain walls in $FeSn_2$. **c,** A picture of single crystals of $FeSn_2$. **d,** Temperature dependence of magnetization measured with $\mu_0 H = 0.1$ T. The insets show schematic top views along the *c*-axis of the magnetic structures in the collinear and canted AFM phases. **e,** Temperature dependence of magnetization measured with $\mu_0 H = 0.1$ T for three magnetic field directions ([100], [010], and [001]).

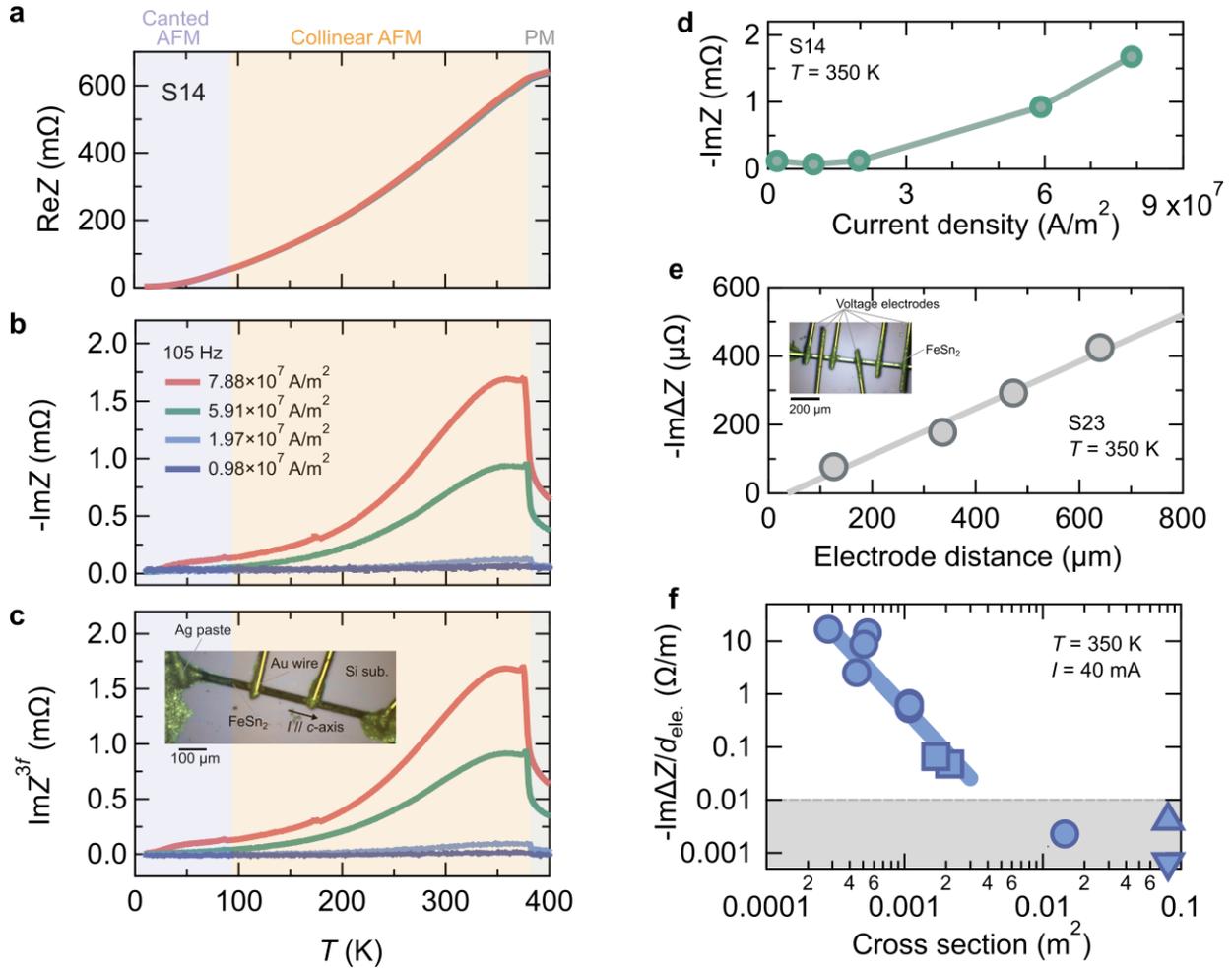

**Figure 3 | Emergent electric field in FeSn$_2$. a-c,** Temperature dependence of the real part (a) and the imaginary part (b) of the complex impedance and the imaginary part of the third-harmonic complex impedance (c) measured at various current densities under zero magnetic field. The inset shows an optical image of the sample. **d,** Current-density dependence of the imaginary part of the complex impedance at 350 K. **e,** Magnitude of the imaginary part of the background-subtracted complex impedance (Im$\Delta Z$) as a function of the distance between the voltage electrodes at 350 K. The inset shows an optical image of the sample used in this measurement. **f,** Magnitude of Im$\Delta Z$ normalized by the electrode distance ($d_{\text{ele.}}$) as a function of the cross-sectional area of the samples at 350 K. The gray dashed line represents the detection limit of our system. The circles, squares, and triangle represent the electrode attachment methods. The circles correspond to silver paste (Dupont 6838), the squares to welding, the triangles to silver paste (Dupont 4622), and the inverted triangle to indium solder (see Methods for details).

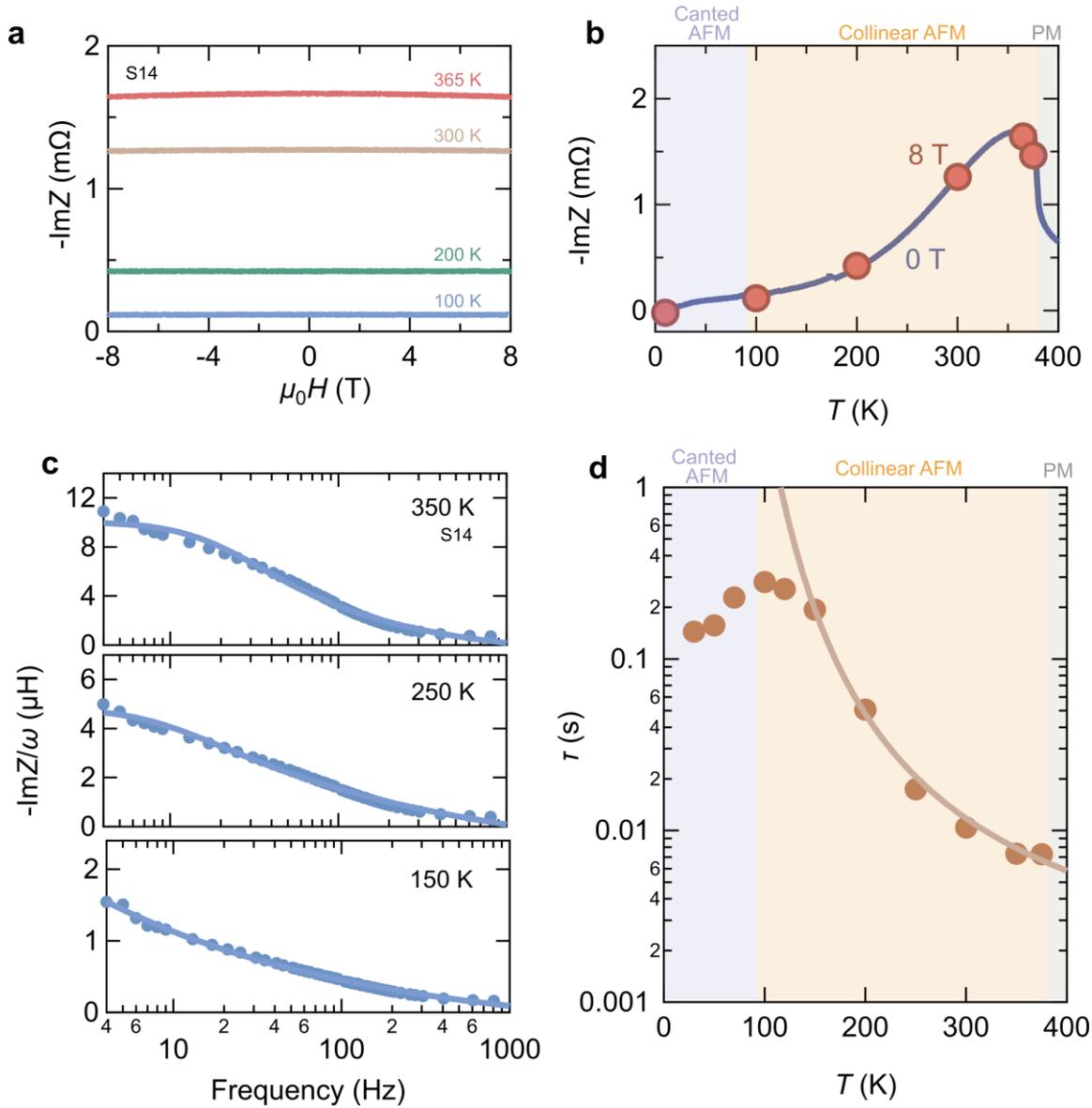

**Figure 4 | Magnetic-field and frequency dependence of emergent electric field. a, b**, Magnetic field dependence of the imaginary part of the complex impedance (Im$Z$) measured with $j = 7.88 \times 10^7$ A/m$^2$. **c**, Frequency dependence of Im$Z$ measured with $j = 7.88 \times 10^7$ A/m$^2$. The solid curves are fits to an extended Debye model (see Methods). **d**, Temperature dependence of the typical relaxation time $\tau$ evaluated from the fits to the extended Debye model. The solid curve is a fit to an Arrhenius equation.

Supplementary Information for

# Emergent electric field induced by current-driven domain wall motion in a room-temperature antiferromagnet FeSn$_2$

## Theoretical calculations

Under an ac electric current density $j(t) = j_\omega e^{i\omega t}$ applied along the $x$ axis, we assume for the Néel vector $\boldsymbol{l} = (\sin\theta\cos\phi, \sin\theta\sin\phi, \cos\theta)$ the following one-dimensional DW profile

$$\theta(t,x) = 2\tan^{-1}\left[\exp\left(q\frac{x-X(t)}{\lambda}\right)\right], \quad \phi(t,x) = \Psi(t), \tag{1}$$

where $X(t)$ and $\Psi(t)$ are the collective coordinates describing the DW position and the angle of the DW plane, respectively, $\lambda$ is the DW width parameter, and $q = \frac{1}{\pi}\int dx \frac{\partial\theta}{\partial x} = \pm 1$. Substituting Eq. (1) in a general equation of motion for $\boldsymbol{n}$ that takes into account the temporal variation of the current[1,2], one obtains

$$\frac{1}{\gamma H_E}\frac{d^2 X}{dt^2} + \alpha\frac{dX}{dt} = \beta u + \frac{1}{\gamma H_E}\frac{\partial u}{\partial t} + F_{\text{pin}}, \tag{2}$$

$$\frac{1}{\gamma H_E}\frac{d^2\Psi}{dt^2} + \alpha\frac{d\Psi}{dt} = 0. \tag{3}$$

Here, $\gamma$ is the gyromagnetic ratio, $\alpha$ is the dimensionless damping constant, $H_E$ is the antiferromagnetic exchange coupling field, $F_{\text{pin}}$ is the force due to pinning potentials for the DW, $\beta$ is the so called nonadiabatic parameter, and $u$ is a parameter that is proportional to the electric current density and in the unit of velocity. In the limit where the exchange coupling between the conduction electrons and the local magnetizations is dominant over the inter-sublattice electron hopping energy, $u$ is given by $u(t) = -\frac{\gamma\hbar p}{2e\mu_0 M_S}j(t)$, with $e(>0)$ the elementary electric charge, $M_S$ the saturation magnetization of each sublattice, and $p$ the spin polarization of the conduction electrons in each sublattice. More generally, it has been pointed out by a microscopic model calculation that $u$ can be significantly enhanced near the antiferromagnetic gap edge[3]. Note that the equations of motion for $X$ and $\Psi$ are decoupled and the current does not drive the dynamics of $\Psi$; therefore we can focus only on the emergent electric field generated by the translational motion of the DW.

We first consider a harmonic pinning potential with $F_{\text{pin}} = -\omega_{\text{pin}}^h X$, with $\omega_{\text{pin}}^h$ a constant characterizing the potential. In the low frequency limit where $\omega \ll \sqrt{\gamma H_E \omega_{\text{pin}}^h}$ and $\alpha\omega \ll \omega_{\text{pin}}^h$ are simultaneously satisfied, Eq. (2) can be approximately solved as

$$X(t) = \frac{\beta}{\omega_{\text{pin}}^h} u(t) \equiv \frac{\beta}{\omega_{\text{pin}}^h} u_\omega e^{i\omega t}. \qquad (4)$$

Using this into the expression for the DW-induced electric voltage (see the main text), one obtains

$$V(t) = -\frac{\hbar p}{e\lambda} \frac{\beta^2}{\omega_{\text{pin}}^h} i\omega u_\omega e^{i\omega t}. \qquad (5)$$

$V$ is now proportional to the time derivative of the current, because under the condition considered in this paragraph the DW velocity takes a pure imaginary value, i.e., perfectly out of phase with the ac current. We therefore define the emergent inductance (the inductance due to $V$) in the standard way as

$$L = -\left[\frac{d(S_c j)}{dt}\right]^{-1} V = -\left(\frac{\hbar p}{e}\right)^2 \frac{\beta^2}{\omega_{\text{pin}}^h} \frac{\gamma}{2\mu_0 M_S \lambda S_c}, \qquad (6)$$

where $S_c$ is the cross-sectional area of the sample.

Next, we show that introducing anharmonicity in the pinning potential leads to higher harmonic components of $V$. For the sake of a proof-of-concept type of demonstration, we employ a specific anharmonic pinning potential $F_{\text{pin}} = -\omega_{\text{pin}}^{\text{ah}} \frac{\sin(2kX)}{k}$, where $\omega_{\text{pin}}^{\text{ah}}$ and $k$ are constants characterizing, respectively, the amplitude and the spatial period of the potential. We numerically simulate the DW dynamics, Eq. (2), and the Fourier series $V^n = \frac{\omega}{\pi}\int_0^{2\pi/\omega} dt\, V(t) e^{-in\omega t}$ of the resultant electric voltage. In Supplementary Fig. 1, we show the results for $-\text{Im}[\tilde{V}^{n=1,3,5}]$ as functions of $j_\omega$ and $\omega$, where $\tilde{V}^n$ is the normalized n-th component $\tilde{V}^n = V^n(j_\omega, \omega)/V^1(j_\omega = 10^8\,\text{Am}^{-2}, \omega/2\pi = 0.1\,\text{kHz})$. It can be seen that the higher harmonic components are more pronounced with larger $j_\omega$ and lower $\omega$, as the DW travels greater distance during the time period $2\pi/\omega$. Each component can take either a positive or a negative sign depending on the current amplitude and frequency. The set of parameter values used is $\mu_0 M_S = 1$ T, $\mu_0 H_E = 25.13$ T, $\alpha = 0.01$, $\beta = 0.02$, $p = 0.4$, $\lambda = 10$ nm, $\omega_{\text{pin}}^{\text{ah}} = 1$ s$^{-1}$, and $k =$

$5 \times 10^4$ m$^{-1}$. For the calculation in Fig. 1c, we employed the same shape of the anharmonic pinning potential with $\omega_{\text{pin}}^{\text{ah}} = 100$ s$^{-1}$ and $k = 5 \times 10^4$ m$^{-1}$, while the other parameters are the same as here.

We note that we have ignored corrections, due to the canting of the sublattice magnetizations, to the DW dynamics and the emergent electric field. We have also focused on low frequency regimes where the $\partial u/\partial t$ term in Eq. (2) is negligible. These effects are largely irrelevant to the analysis of our present experiment and will be discussed elsewhere.

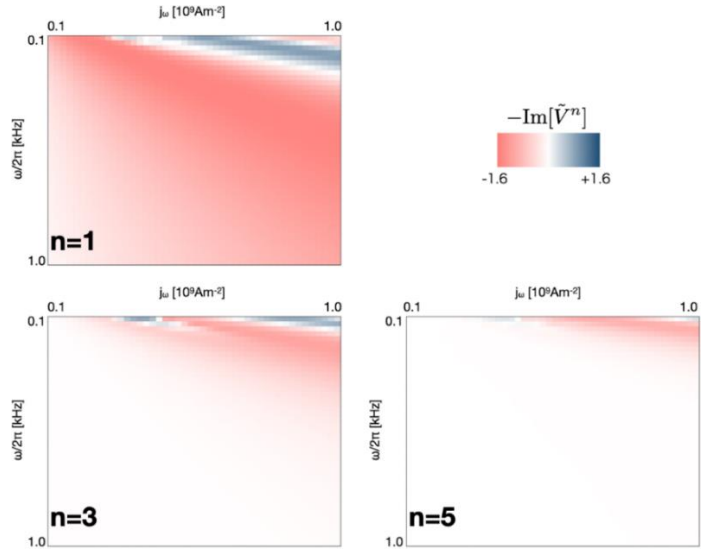

**Supplementary Fig. 1** | $-\text{Im}[\tilde{V}^{n=1,3,5}]$ as functions of $j_\omega$ and $\omega$, where $\tilde{V}^n$ is the normalized n-th component.

## Magnetic and transport properties of FeSn$_2$

In Supplementary Fig. 2, we show the magnetic and transport properties of FeSn2 (S10), with dimensions of 0.31×0.26×2 mm$^3$. The magnetization linearly increases with increasing magnetic field (Supplementary Fig 2a), which is consistent with the formation of an antiferromagnetic order.

Supplementarily Figs. 2b-d show the transport properties of FeSn$_2$. The resistivity decreases with decreasing temperature, indicating good metallic conduction with a residual resistance ratio greater than 149. FeSn$_2$ exhibits a large magnetoresistance at low temperatures (Supplementary Figs. 2c and d). A similar large magnetoresistance has been observed in FeGe$_2$, a sister compound of FeSn$_2$, because of the Lorentz force acting on the high-mobility carriers[4,5]. Thus, the large magnetoresistance in FeSn$_2$ is likely to stem from the Lorentz force acting on high-mobility carriers. As shown in Supplementary Fig. 2e, the Hall resistivity is linear with respect to the magnetic field. Since the Hall resistivity in AFM is the sum of the normal Hall effect and the anomalous Hall effect, this field profile suggests that the magnetization linearly rises with increasing magnetic field. In other words, no spin-flip transition occurs below 9 T.

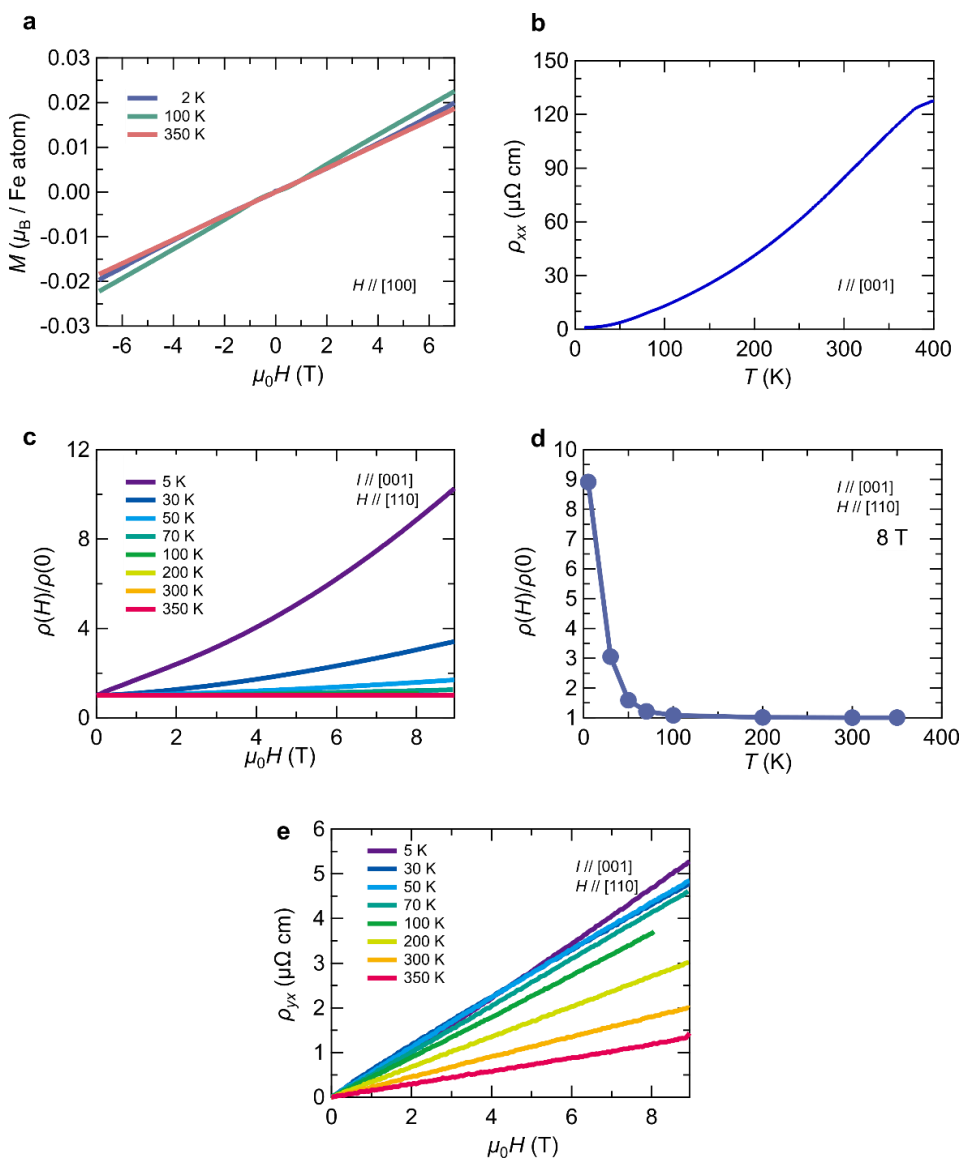

**Supplementary Fig. 2 | a**, Magnetic field dependence of magnetization. **b**, Temperature dependence of resistivity **c** and **d**, Magnetic field (c) and temperature (d) dependence of resistivity. **e,** Magnetic field dependence of Hall resistivity.

# Sample dependence of emergent electric field

In Supplementary Fig. 3, we show the temperature dependence of Re$Z$, Im$Z$, and Im$\Delta Z$ normalized by the electrode distance for the samples with various cross-sectional areas. Supplementary Fig. 4 shows the temperature dependence of Re$Z$ and Im$Z$ at different electrode distances.

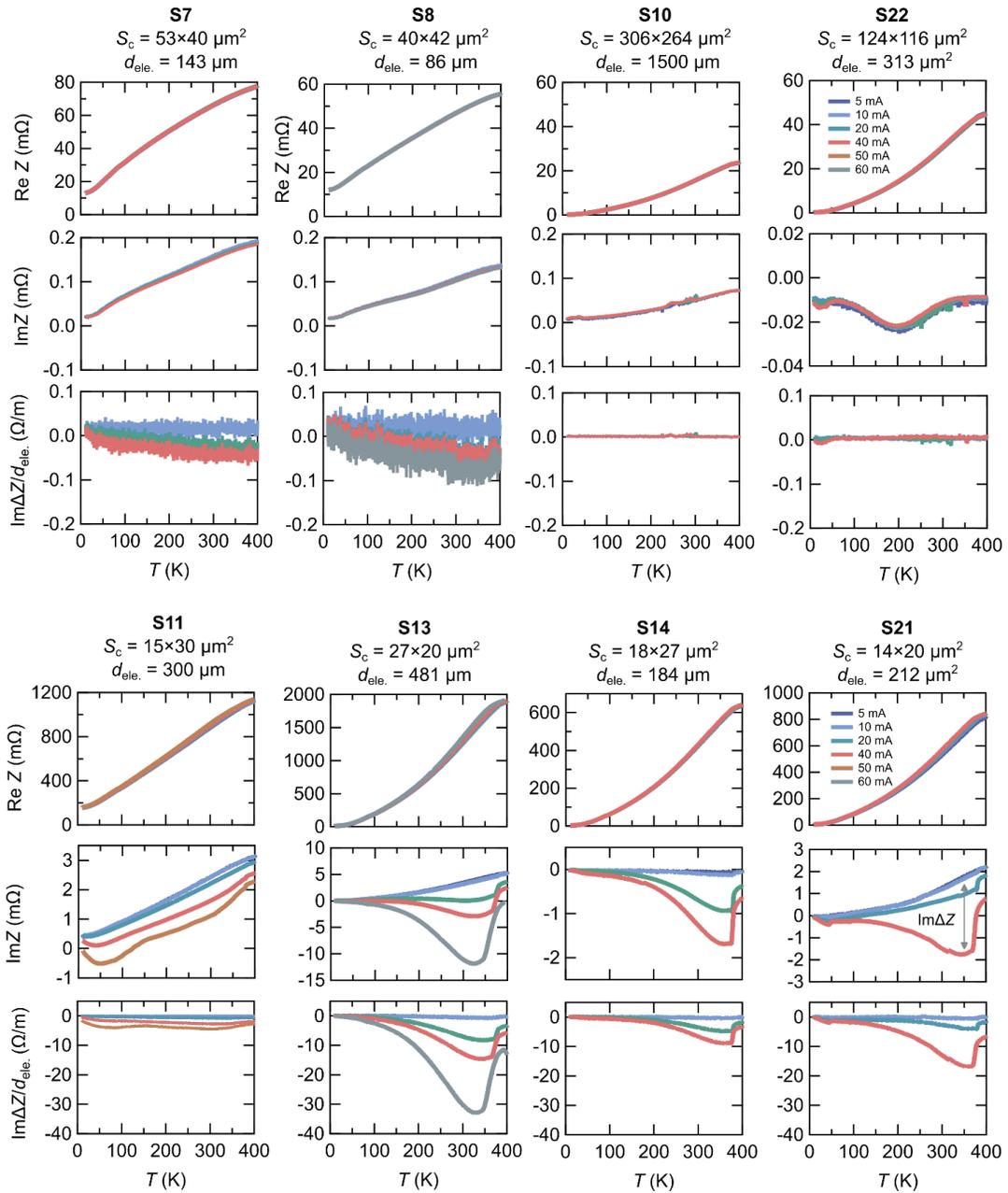

**Supplementary Fig. 3** | Temperature dependence of Re$Z$, Im$Z$, and Im$\Delta Z$ normalized by the electrode distance for the samples with various cross-sectional areas.

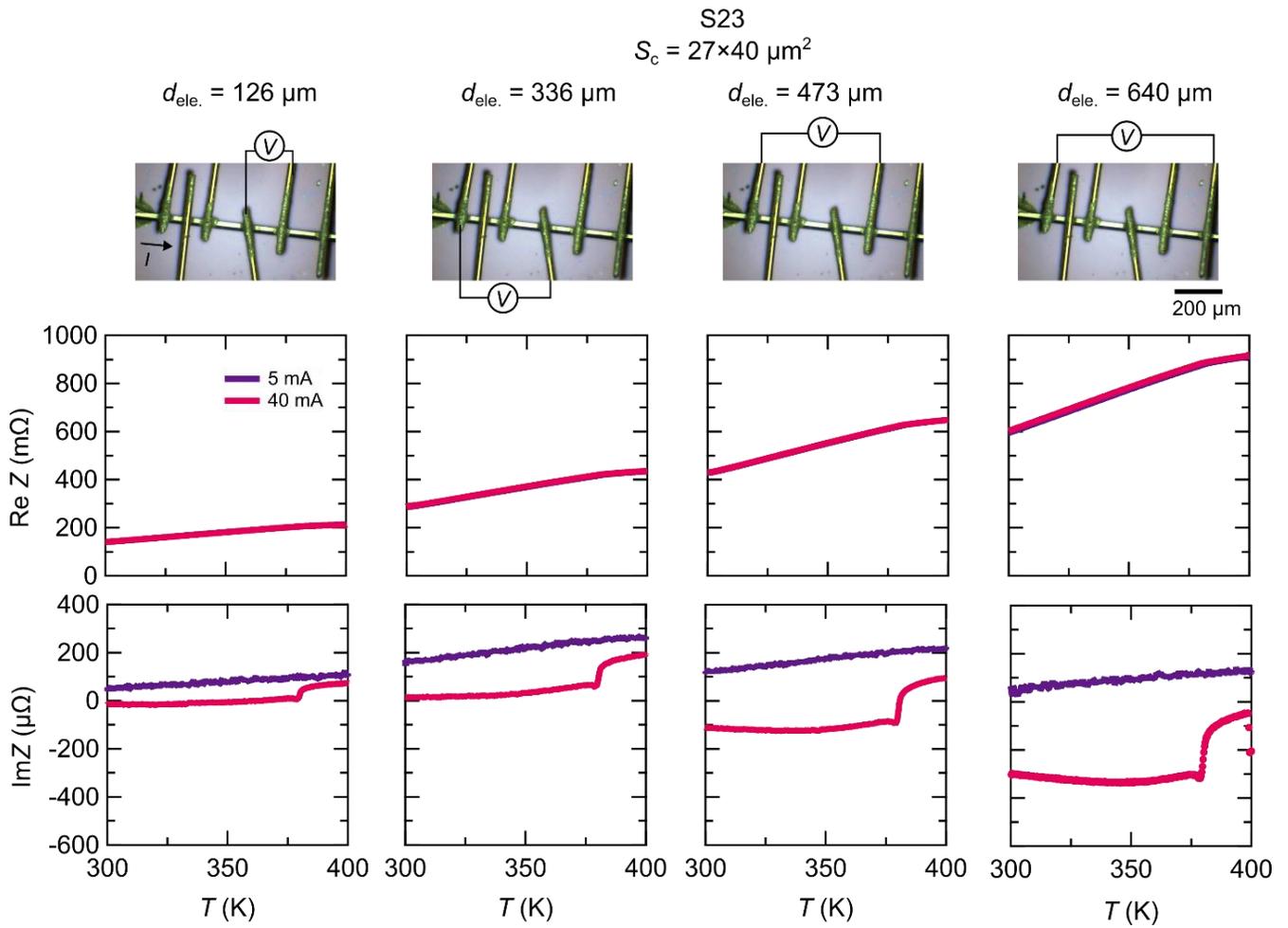

**Supplementary Fig. 4** | Temperature dependence of Re$Z$ and Im$Z$ for different electrode distances measured at a frequency of $f$ = 105 Hz and a currents of $I$ = 5 and 40 mA.